\documentclass[twocolumn,showpacs,preprintnumbers, prc,superscriptaddress]{revtex4}
\usepackage{epsfig}
\usepackage{graphicx}
\usepackage{amsmath,amssymb,amsfonts}
\usepackage{array}
\usepackage{url}
\usepackage{hyperref}
\usepackage{lineno}
\usepackage{xspace}
\usepackage[usenames,dvipsnames]{color}
\newcommand{\sqsn}{\mbox{$\sqrt{s_{_{NN}}}$}\xspace}
\newcommand{\bef}{\begin{figure}}
\newcommand{\eef}{\end{figure}}
\newcommand{\bc}{\begin{center}}
\newcommand{\ec}{\end{center}}

\newcommand{\be}{\begin{equation}}
\newcommand{\ee}{\end{equation}}
\newcommand{\bea}{\begin{eqnarray}}
\newcommand{\eea}{\end{eqnarray}}

\begin{document}
\title{Correlations of conserved number mixed susceptibilities in a hadron 
resonance gas model}
\author{D.~K.~Mishra}
\email {dkmishra@rcf.rhic.bnl.gov}
\affiliation{Nuclear Physics Division, Bhabha Atomic Research Center, Mumbai 
400085, India}
\author{P.~K.~Netrakanti}
\email {pawannetrakanti@gmail.com}
\affiliation{Nuclear Physics Division, Bhabha Atomic Research Center, Mumbai 
400085, India}
\author{Bedangadas Mohanty}
\email {bedanga@niser.ac.in}
\affiliation{National Institute of Science Education and Research, Jatni 
752050, India}

\begin{abstract}
The ratios of off-diagonal and diagonal susceptibilities of conserved charges 
are studied using a hadron resonance gas (HRG) model with an emphasis towards 
providing a proper baseline for comparison to the corresponding future 
experimental measurements. We have studied the effect of kinematic acceptances, 
transverse momentum ($p_T$) and pseudorapidity ($\eta$), and different charged 
states on the ratios of the calculated susceptibilities. We find that the effect 
of $p_T$ and $\eta$ acceptance on the ratio of the susceptibilities are small 
relative to their dependence on the beam energy or the charged states of the 
used particles. We also present a HRG based calculation 
for various combinations of cumulant ratios of protons and pions, recently 
proposed as robust observables (with no theoretical uncertainties) for critical 
point search in the experiments. These results which increase as a function of 
collision energy will provide a better baseline for non-critical point physics 
compared to Poisson expectation. 

\end{abstract}

\maketitle

\section{Introduction}
\label{intro} 
In recent years, beam energy scan (BES) program carried out at Relativistic 
Heavy-Ion Collider (RHIC) has drawn much attention with an aim to explore the 
quantum chromodynamics (QCD) phase diagram at non-zero temperature ($T$) and 
baryon chemical potential 
($\mu_{B}$)~\cite{Stephanov:1998dy,Alford:1997zt,Aoki:2006we}. One 
of the key observable is related to the event-by-event distribution of 
conserved charge numbers such as: net-baryon, net-electric charge and 
net-strangeness~\cite{Asakawa:2000wh,Asakawa:2009aj,Stephanov:2004wx}. Current 
experiments at RHIC have reported the measurements for various order moments of 
net-proton and net-charge event-by-event distributions using the BES phase-I 
data~\cite{Aggarwal:2010wy,Adamczyk:2013dal,Adamczyk:2014fia,Adare:2015aqk}. 
From the net-proton measurement, deviations are observed for \sqsn $\le$ 27 GeV 
for ratios of higher order cumulants compared to expectation from a Hadron 
Resonance Gas (HRG) model and Skellam statistics~\cite{Adamczyk:2013dal}. 
However, from the experimental side, there is no conclusive evidence about the 
location of the QCD critical end point (CEP). This has prompted to have high 
statistics BES-II program scheduled for 2018-19 with the focus on taking data at 
lower beam energies (below 39 GeV). So far, the reported experimental 
measurements are based on higher moments of either net-proton or net-charge 
multiplicity distribution, which are related to baryon number or electric charge 
susceptibilities, respectively~\cite{Aggarwal:2010wy,Gupta:2011wh}. 

Probes of phase structure of QCD should also include correlations
among different conserved charges~\cite{Bazavov:2012jq}. They are not only 
expected to vary in some characteristic manner between low and high temperature 
phases of QCD but also provide insight into applicability of the HRG model. It 
has been suggested that, the ratio of strange to non-strange susceptibilities 
calculated in lattice QCD can give insight about Quark Gluon Plasma (QGP) 
phase~\cite{Gavai:2002kq}. The correlation between baryon number ($B$) and 
electric charge ($Q$) ($\chi_{BQ}$) shows a variation with the temperature which 
are correlated with the changes in relevant degrees of 
freedom~\cite{Bazavov:2012jq}. At lower temperature $\chi_{BQ}$ is dominated by 
contributions from protons and anti-protons, whereas in the high temperature 
limit of (2+1) flavor QCD, their values are due to quark degrees of freedom. 
Similarly, correlations of baryons with strangeness ($\chi_{BS}$) and electric 
charge with strangeness ($\chi_{QS}$) are sensitive to the strangeness degrees 
of freedom of the system~\cite{Koch:2005vg,Majumder:2006nq}. At high 
temperature limit, in which the basic degrees of freedom are quarks and gluons, 
the strangeness is carried by the $s$ and $\bar s$ quarks that later carry the 
baryon number in strict proportion to their strangeness number 
($B=-\frac{1}{3}S$ ) which makes the $B$ and $S$ strongly 
correlated~\cite{Koch:2005vg}. On the other hand, in a hadron gas the 
strangeness is mostly carried by kaons with $B \approx 0$, which makes the 
correlation between strangeness and baryon number very small. Hence, observables 
such as $\chi_{QS}$, $\chi_{BQ}$ and $\chi_{BS}$ give information about the QCD 
phase structure~\cite{Bazavov:2012jq}.

Further, combinations of diagonal and off-diagonal quark number susceptibilities 
can be used to obtain the fluctuations of conserved 
charges~\cite{Gavai:2005yk,Bazavov:2012vg,Borsanyi:2011sw}. Diagonal 
susceptibilities measure the quark number density to changes in $T$ and $\mu_B$. 
At high temperature region diagonal susceptibilities show large values and 
expected to approach the ideal gas limit~\cite{Borsanyi:2011sw}. Whereas in the 
low temperature region, the quarks are confined, the diagonal susceptibilities 
are expected to have small values. The off-diagonal susceptibilities in turn can 
be used to explore the degree of correlation between different charge and baryon 
numbers that carry different 
flavors~\cite{Gavai:2005yk,Bazavov:2012vg,Borsanyi:2011sw}. Using the ratio of 
susceptibilities, $\chi_{BS}/\chi_S$ and $\chi_{QS}/\chi_S$, it has been observed 
that, above critical temperature ($T_c$), the flavor carrying sector of QCD is 
consistent with deconfined plasma of weakly interacting quarks and antiquarks. 
Above 1.5$T_c$, the flavor carrying degrees of freedom are quark-like 
quasi-particles. Below $T_c$ the behavior of the above susceptibility ratios is 
consistent with that of hadron resonance gas~\cite{Majumder:2006nq}.

In addition, a landmark point in the QCD phase diagram is the 
CEP~\cite{Stephanov:1999zu,Fodor:2004nz}. As one approaches the CEP in the 
phase diagram, there will be large fluctuations in the event-by-event produced 
particle numbers~\cite{Athanasiou:2010kw}. These fluctuations are quantified by
measuring various order moments of the event-by-event particle number 
distribution. These moments are related to some power of correlation length 
($\xi$) of the 
system~\cite{Stephanov:1999zu,Stephanov:2008qz,Gavai:2010zn,Cheng:2008zh}. 
Higher moments of the distributions are related to the higher power of $\xi$ 
which makes them more sensitive to CEP 
effects~\cite{Stephanov:2011pb,Asakawa:2009aj}. Specifically it has been
recently suggested to look for mixed pion-proton cumulants as a signature for 
CEP~\cite{Athanasiou:2010kw}. The paper suggests to construct five specific 
mixed pion-proton cumulants from which when non-CEP contributions are 
subtracted will yield a value of unity if the system experiences critical 
phenomena. Such ratios have no theoretical uncertainties arising from 
uncertainty in the values of the QCD based parameters. These ratios are claimed 
to provide strong evidence needed for the discovery of the QCD critical point.

Keeping in mind above aspects and the importance of having a proper baseline 
for the experimental measures, we discuss in the present paper, mixed 
susceptibilities (between electric charge, baryon number, and strangeness) and 
various order pion-proton cumulants within the frame work of a HRG 
model~\cite{BraunMunzinger:2003zd,Cleymans:2005xv,Andronic:2011yq}. The 
calculations presented here also keeps in mind the actual experimental situation 
in terms of acceptances (in momentum and rapidity) and measurable 
quantities~\cite{Garg:2013ata,Karsch:2015zna}. 

The paper is organized as follows. In the following section, we discuss the HRG 
model used in this study. In Section~\ref{sec:acceptance}, the results for 
various mixed cumulants are presented for different kinematic acceptance and 
charge states. In Section~\ref{sec:baselineratios}, we present the proper 
baseline for critical fluctuation. Finally in Section~\ref{sec:summary}, we 
summarize our findings and mention about the implications of this work in 
current experimental measurements in high energy heavy-ion collisions.

\section{Mixed susceptibilities in hadron resonance gas model}
The partition function ($Z$) in the HRG model include all the degrees of freedom 
of confined, strongly interacting matter and implicitly contains all the 
interactions that result in resonance formation~\cite{Karsch:2010ck}. The 
logarithm of the partition function is given by 
\begin{equation}
ln Z (T,\mu,V) = \sum_M ln Z_i(T,\mu_i,V) + \sum_B ln Z_i(T,\mu_i,V)
\end{equation}
where 
\begin{equation}
 ln Z_i(T,\mu_i,V) = \pm \frac{Vg_i}{2\pi^2}\int d^3p ~ln\big\{1\pm 
 \mathrm{exp}[(\mu_i-E)/T]\big\},
\end{equation}
$T$ is the temperature, $V$ is the volume of the system, $\mu_{i}$ is the 
chemical potential and $g_{i}$ is the degeneracy factor of the $i^{th}$ 
particle. The total chemical potential $\mu_i$ = $B_i\mu_{B}$ + $Q_i\mu_{Q}$ + 
$S_i\mu_{S}$, where $B_i$, $Q_i$ and $S_i$ are the baryon, electric charge  and 
strangeness number of the $i^{th}$ particle, with corresponding chemical 
potentials $\mu_{B}$, $\mu_{Q}$ and $\mu_{S}$, respectively. The $+ve$ and 
$-ve$ signs are for baryons and mesons respectively. The thermodynamic pressure 
($P$) can then be obtained from the logarithm of partition function in the limit 
of large volume as:
\begin{equation}
 P(T,\mu_i,V) = \frac{T}{V} ln Z_i = 
\pm\frac{T g_{i}}{2\pi^2}\int d^3{k}\ln{\big\{1\pm\exp(\mu_i-E)/T}\big \}    
\end{equation}
In a static fireball, a particle of mass $m$ with $p_T$, $\eta$ and $\phi$ as 
the transverse momentum, pseudo-rapidity and azimuthal angle,
respectively, the 
volume element ($d^3p$) and energy ($E$) of the particle can be written as $d^3p 
= p_Tm_Tcosh\eta dp_Td\eta d\phi$ and $E = m_Tcosh\eta$, where $m_T$ is the 
transverse mass $= \sqrt{m^2 + p_T^2}$. The experimental acceptances can be 
applied by considering the corresponding ranges in $p_T$, $\eta$ and 
$\phi$~\cite{Garg:2013ata}. The fluctuations of the conserved numbers are 
obtained from the derivative of the thermodynamic pressure with respect to the 
corresponding chemical potentials $\mu_B, \mu_Q$ or $\mu_S$. The $n$-th order 
generalized susceptibilities ($\chi_x$), where $x$ represents baryon, 
electric charge or strangeness indices, can be expressed as;
\begin{equation}
 \chi_x^n = \frac{d^n[P(T,\mu)/T^4]}{d(\mu_x/T)^n}    
\end{equation}
For mesons $\chi_x$ can be expressed as 
\begin{eqnarray}
 \chi_{x,meson}^{(n)}=\frac{X^n}{VT^3}
\int{d^{3}p}\sum_{k=0}^{\infty}(k+1)^{n-1} \nonumber \\  
\times ~ \mathrm {exp}\bigg\{\frac {-(k+1)E } {T}\bigg\} {\mathrm{exp}\bigg\{ 
\frac{(k+1)\mu}{T}
\bigg\}}.
\label{eq:susc_mes}
 \end{eqnarray}
 and for baryons,
\begin{eqnarray}
\chi_{x,baryon}^{(n)}=\frac{X^n}{VT^3} \int{d^{3}p}\sum_{k=0}^{\infty}{(-1)^k}
(k+1)^{n-1} \nonumber \\ 
\times ~ \mathrm{exp}\bigg\{\frac{-(k+1)E} {T}\bigg\} 
{\mathrm{exp}\bigg\{\frac{(k+1)\mu}
{T}\bigg\}},
\label{eq:susc_bar}
\end{eqnarray}
where $X$ represents either $B_i$, $Q_i$ or $S_i$ of the $i$-th particle.
The total generalized susceptibilities will be the sum of susceptibility of 
mesons and baryons as $\chi_x^{n} = \sum \chi_{x,mesons}^n + \sum 
\chi_{x,baryons}^n$. Further, the mixed susceptibilities of the correlated 
conserved charges can be obtained by taking the derivative of the pressure with 
respect to different chemical potentials for conserved quantities $X$ and $Y$,
\begin{equation}
\chi_{xy}^{(n,m)} = \frac{d^{n+m}[P(T,\mu)/T^4]}{d(\mu_x/T)^n d(\mu_y/T)^m}
\end{equation}
where $m$ and $n$ correspond to the different order of derivatives, $x$ and $y$ 
are baryon--electric charge, baryon--strangeness or electric charge--strangeness 
mixed indices and $X, Y$ are combinations of either $B, Q$ or $S$. Hence, 
Eq.~\ref{eq:susc_mes} and ~\ref{eq:susc_bar} can be written as:

\begin{eqnarray}
 \chi_{xy,meson}^{(n,m)}&=&\frac{X^nY^m}{VT^3}
\int{d^{3}p}\sum_{k=0}^{\infty}(k+1)^{n-1}(k+1)^{m-1} \nonumber \\  
&\times&\mathrm{exp}\bigg\{\frac {-(k+1)E } {T}\bigg\} {\mathrm{exp}\bigg\{ 
\frac{(k+1)\mu}{T}\bigg\}},
\label{eq:corrsusc_mes}
 \end{eqnarray}
and
\begin{eqnarray}
\chi_{xy,baryon}^{(n,m)}&=&\frac{X^nY^m}{VT^3} \int{d^{3}p}\sum_{k=0}^{\infty}
{(-1)^k}(k+1)^{n-1}(k+1)^{m-1} \nonumber \\ 
&\times&\mathrm{exp}\bigg\{\frac{-(k+1)E} {T}\bigg\} 
{\mathrm{exp}\bigg\{\frac{(k+1)\mu}{T}\bigg\}}.
\label{eq:corrsusc_bar}
\end{eqnarray}

Using Eq.~\ref{eq:corrsusc_mes} and \ref{eq:corrsusc_bar}, one can calculate the
conserved charge correlations in the HRG model. The freeze-out parameters used 
in the HRG are extracted from a statistical model description of particle 
yields. The parametrization of $\mu_B$ and $T$ as a function collision energies 
(\sqsn) are given in~\cite{Cleymans:2005xv,Karsch:2010ck}. Other effects like
collective flow and resonances on susceptibilities are discussed in 
Ref.~\cite{Garg:2013ata}.
\begin{figure}[t]
 \begin{center}
\includegraphics[width=0.5\textwidth]{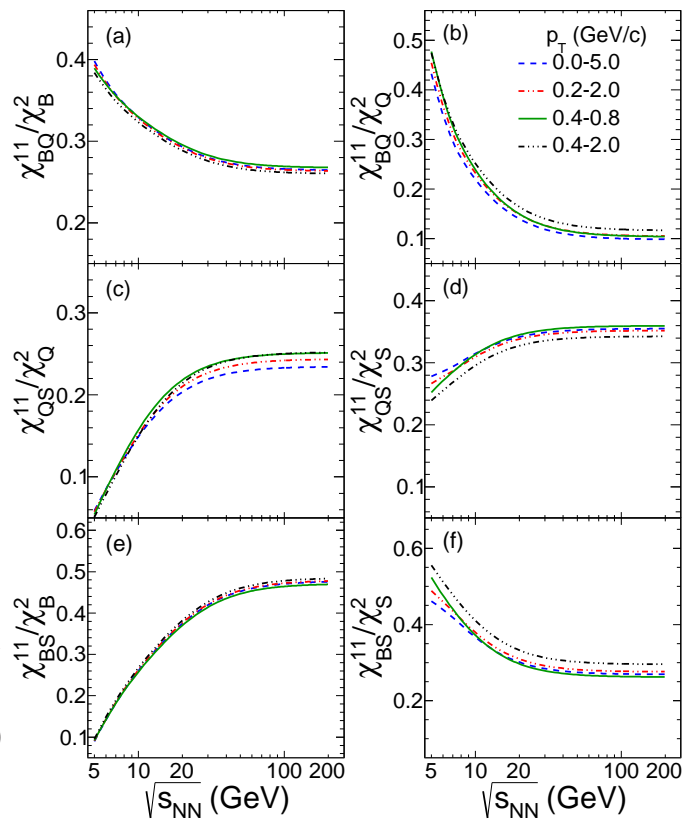}
 \caption{Ratios of diagonal susceptibility to second order susceptibility 
($\chi_{xy}^{11}/\chi_{x,~y}^2$ ) as a function of \sqsn within $|\eta| \le$ 0.5 
from the HRG model, where $x$ and $y$ stand for either $B$, $Q$ or $S$. Panels 
(a) and (b) shows the diagonal susceptibility ratios for charged baryons. Panels 
(c) and (d) shows the diagonal susceptibility ratios for charged particles with 
at least one strange quark. Panels (e) and (f) shows the diagonal 
susceptibility ratios for baryons with strange quark. The ratios are also 
compared for various $p_{T}$ acceptance ranges.}
 \label{fig:netBQ11_pt}
 \end{center}
 \end{figure}

\section{Acceptance effect on correlated susceptibilities}
\label{sec:acceptance}
The phase structure of QCD at finite temperature and baryon chemical
potential can be probed using correlation 
among different conserved charges. There are significant changes of these 
correlations in the crossover region between the low and high temperature phases 
of QCD~\cite{Bazavov:2012jq}. As one goes from low to high temperature phase, it 
is expected that there is change in correlations between baryon number and 
electric charge ($\chi_{BQ}$), between baryon number and strangeness 
number ($\chi_{BS}$), also between strangeness number and electric 
charge ($\chi_{QS}$). In the following subsections, we discuss the effect of 
kinematic acceptances in the various mixed susceptibilities of the conserved 
charges. In the present study we have considered $\chi_{BQ}^{11}/\chi_B^2$, 
$\chi_{BQ}^{11}/\chi_Q^2$, $\chi_{QS}^{11}/\chi_Q^2$, 
$\chi_{QS}^{11}/\chi_S^2$, $\chi_{BS}^{11}/\chi_B^2$ and 
$\chi_{BS}^{11}/\chi_S^2$, ratios. The numerator in the first two ratios 
($\chi_{BQ}^{11}/\chi_B^2$ and $\chi_{BQ}^{11}/\chi_Q^2$) receive contributions 
only from charged baryons and anti-baryons (i.e. particles starting from protons 
to all higher mass charged baryons), where as the denominator receives 
contributions from all baryons or all charge particles. In case of other two 
ratios ($\chi_{QS}^{11}/\chi_Q^2$ and $\chi_{QS}^{11}/\chi_S^2$), the 
numerators receive contributions from charged strange and anti-strange 
particles, with kaon being the first charged strange particle, where as 
denominators receive contributions from all charged or strange particles, 
respectively. Similarly, in case of ratios ($\chi_{BS}^{11}/\chi_B^2$ and 
$\chi_{BS}^{11}/\chi_S^2$), the numerators receive contributions from strange 
baryons (anti-baryons), with particle starting from $\Lambda^0$ and higher 
masses of strange baryons, where as denominators receive contributions from all 
baryons or strange particles. It is to be noted that, in real experimental 
situation, it may be difficult to measure the strange baryons on an 
event-by-event basis and calculate their cumulants for studying the 
fluctuations. Particularly, at lower collision energies, where the strange 
baryon production is very small, the above observable will be difficult to 
measure experimentally.

\subsection{Effect of $p_T$ acceptance}
Figure~\ref{fig:netBQ11_pt} shows the variation of $\chi_{BQ}^{11}/\chi_B^2$, 
$\chi_{BQ}^{11}/\chi_Q^2$, $\chi_{QS}^{11}/\chi_Q^2$, 
$\chi_{QS}^{11}/\chi_S^2$, $\chi_{BS}^{11}/\chi_B^2$, and 
$\chi_{BS}^{11}/\chi_S^2$ ratios as a function of \sqsn for different $p_T$ 
acceptances. There is small $p_T$ dependence of the above ratios. However, 
there is a strong energy dependence for $\chi_{BQ}^{11}/\chi_Q^2$, 
$\chi_{QS}^{11}/\chi_Q^2$, and $\chi_{BS}^{11}/\chi_B^2$ ratios.
\begin{figure}[h]
 \begin{center}
\includegraphics[width=0.5\textwidth]{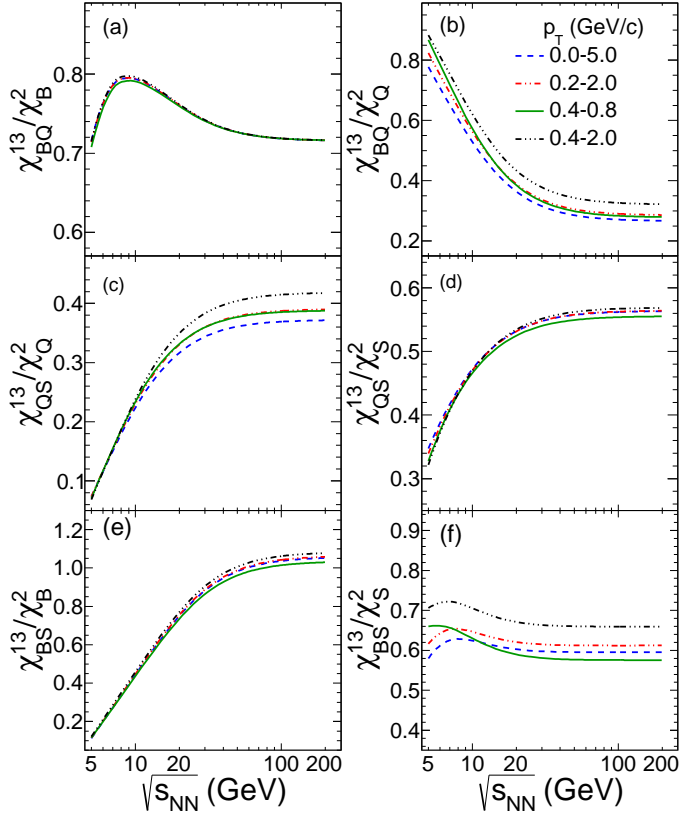}
 \caption{Ratios of off-diagonal susceptibility to second order susceptibility 
($\chi_{xy}^{13}/\chi_{x,~y}^2$ ) as a function of \sqsn within $|\eta| \le$ 
0.5 from HRG, where $x$ and $y$ stand for either $B$, $Q$ or $S$. Panels (a) and 
(b) shows the off-diagonal susceptibility ratios for charged baryons. Panels (c) 
and (d) shows the off-diagonal susceptibility ratios for charged particles with 
at least one strange quark. Panels (e) and (f) shows the off-diagonal 
susceptibility ratios for baryons with strange quark. The ratios are also 
compared for various $p_{T}$ 
acceptance ranges.}
 \label{fig:netBQ13_pt}
 \end{center}
 \end{figure}
\begin{figure}[h]
 \begin{center}
\includegraphics[width=0.5\textwidth]{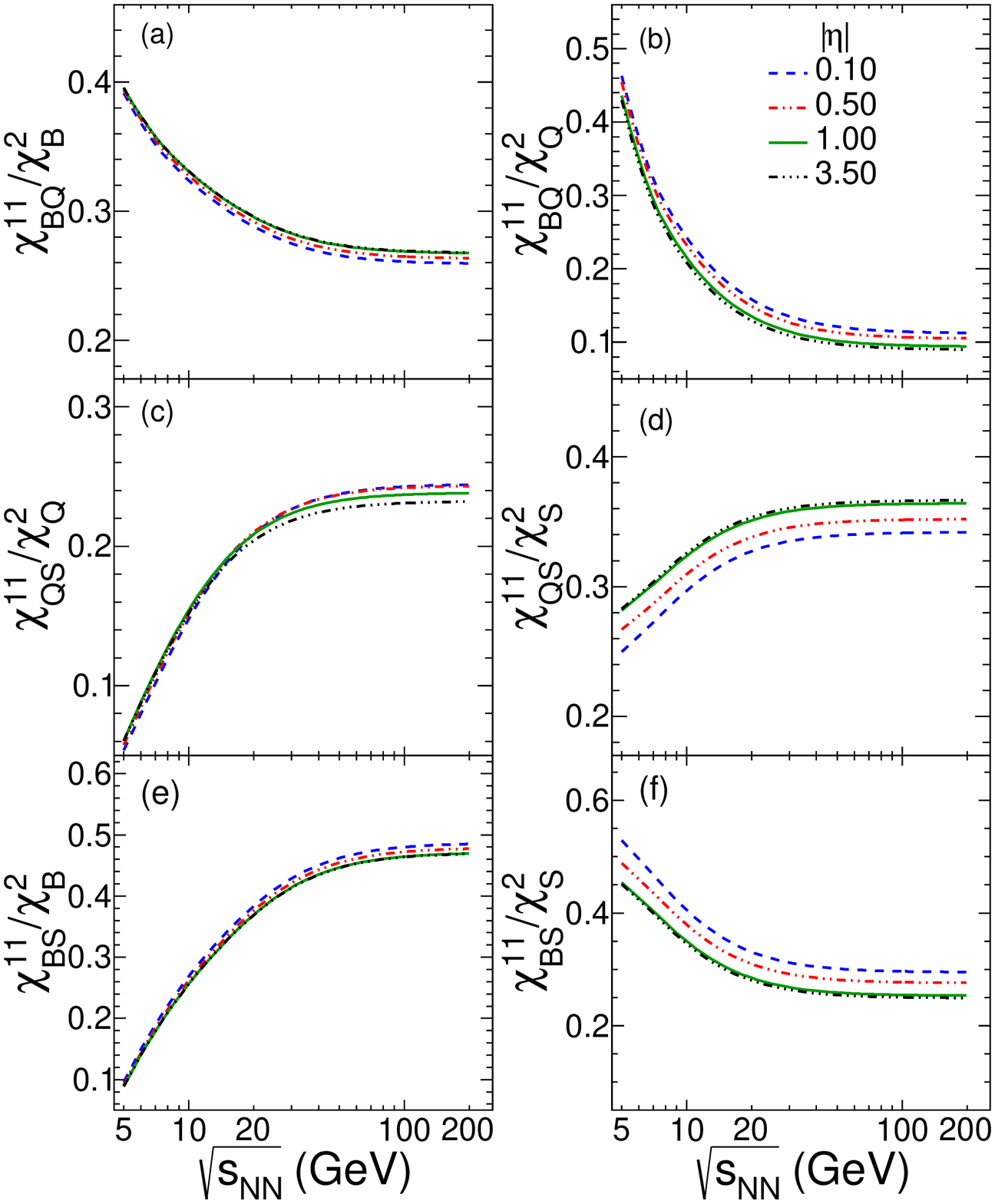}
 \caption{Ratios of diagonal susceptibility to second order susceptibility 
($\chi_{xy}^{11}/\chi_{x,~y}^2$) as a function of \sqsn within 0.2 $\le p_T$ 
GeV/$c \le$ 2.0 from the HRG model, where $x$ and $y$ stand for either $B$, $Q$ 
or $S$. Panels (a) and (b) shows the diagonal susceptibility ratios for charged 
baryons. Panels (c) and (d) shows the diagonal susceptibility ratios for 
charged particles with at least one strange quark. Panels (e) and (f) shows 
the diagonal susceptibility ratios for baryons with strange quark. The ratios 
are also compared for various $\eta$ acceptance ranges.}
 \label{fig:netBQ11_eta}
 \end{center}
 \end{figure}
\begin{figure}[h]
 \begin{center}
\includegraphics[width=0.5\textwidth]{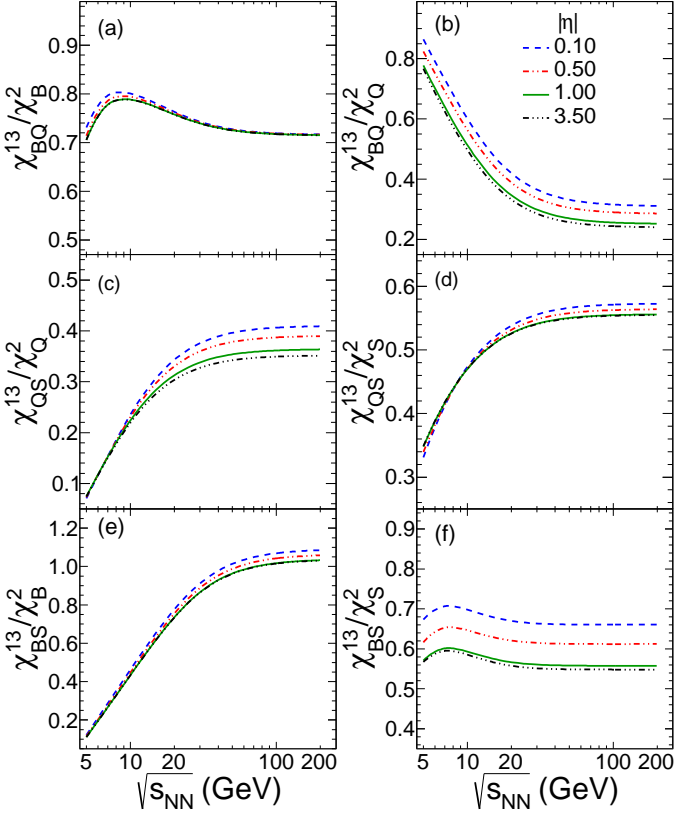}
 \caption{Ratios of off-diagonal susceptibility to second order susceptibility 
($\chi_{xy}^{13}/\chi_{x,~y}^2$) as a function of \sqsn within 0.2 $\le p_T$ 
GeV/$c \le$ 2.0 from the HRG model, where $x$ and $y$ stand for either $B$, $Q$ 
or $S$. Panels (a) and (b) shows the off-diagonal susceptibility ratios for 
charged baryons. Panels (c) and (d) shows the off-diagonal susceptibility 
ratios for charged particles with at least one strange quark. Panels (e) 
and (f) shows the off-diagonal susceptibility ratios for baryons with strange 
quark. The ratios are also compared for various $\eta$ acceptance ranges.}
 \label{fig:netBQ13_eta}
 \end{center}
 \end{figure}
Figure~\ref{fig:netBQ13_pt} shows the corresponding ratios for off-diagonal 
susceptibilities ($\chi_{QS}^{13}/\chi_Q^2$, $\chi_{BQ}^{13}/\chi_Q^2$, 
$\chi_{BQ}^{13}/\chi_B^2$, $\chi_{QS}^{13}/\chi_S^2$, 
$\chi_{BS}^{13}/\chi_B^2$, and $\chi_{BS}^{13}/\chi_S^2$) as a function 
of \sqsn. All the ratios show stronger energy dependence when compared with the 
previous ratios ($\chi^{11}/\chi^2$). There is a visibly $p_T$ dependence 
on $\chi_{BQ}^{13}/\chi_Q^2$, $\chi_{QS}^{13}/\chi_Q^2$, and 
$\chi_{BS}^{13}/\chi_S^2$ ratios for all 
collision energies. The $\chi_{BQ}^{13}/\chi_B^2$, $\chi_{QS}^{13}/\chi_S^2$, 
and $\chi_{BS}^{13}/\chi_B^2$ are least affected by different $p_T$ acceptances.

\subsection{Effect of $\eta$ acceptance}
Figure~\ref{fig:netBQ11_eta} shows the variation of $\chi_{BQ}^{11}/\chi_B^2$, 
$\chi_{BQ}^{11}/\chi_Q^2$, $\chi_{QS}^{11}/\chi_Q^2$, 
$\chi_{QS}^{11}/\chi_S^2$, $\chi_{BS}^{11}/\chi_B^2$, and 
$\chi_{BS}^{11}/\chi_S^2$ ratios as a function of \sqsn for different ranges 
of $\eta$ acceptances. The $p_T$ range for all the particles are considered 
within 0.2 -- 2.0 GeV/$c$. Similar to $p_T$ dependence, there is small $\eta$ 
dependence of the above ratios. However, the $\eta$ dependence is more visible 
in case of $\chi_{QS}^{11}/\chi_S^2$ and $\chi_{BS}^{11}/\chi_S^2$ ratios with 
compared to other three ratios. The $\eta$ 
dependence is even higher in case of higher order mixed susceptibility ratios 
($\chi^{13}/\chi^2$) as a function of collision energies, which are shown 
in Fig.~\ref{fig:netBQ13_eta}. There is no effect of $\eta$ acceptances on the 
$\chi^{13}_{BQ}/\chi^2_B$ ratios. It is to be noted that, the 
baryon-charge correlations ($\chi^{13}_{BQ}/\chi^2_B$) 
in Figs.~\ref{fig:netBQ13_pt} and \ref{fig:netBQ13_eta} show non-monotonic 
behavior as a function of \sqsn could be due to baryon to meson dominated 
freeze-out of the system~\cite{Oeschler:2007bj}. The different energy 
dependence behavior between $\chi_{BQ}^{11}/\chi_B^2$ and 
$\chi_{BQ}^{13}/\chi_B^2$ is primarily due to contribution from 
resonances with higher electric charge (in particular $\Delta^{++}$).

\begin{figure}[h]
 \begin{center}
\includegraphics[width=0.5\textwidth]{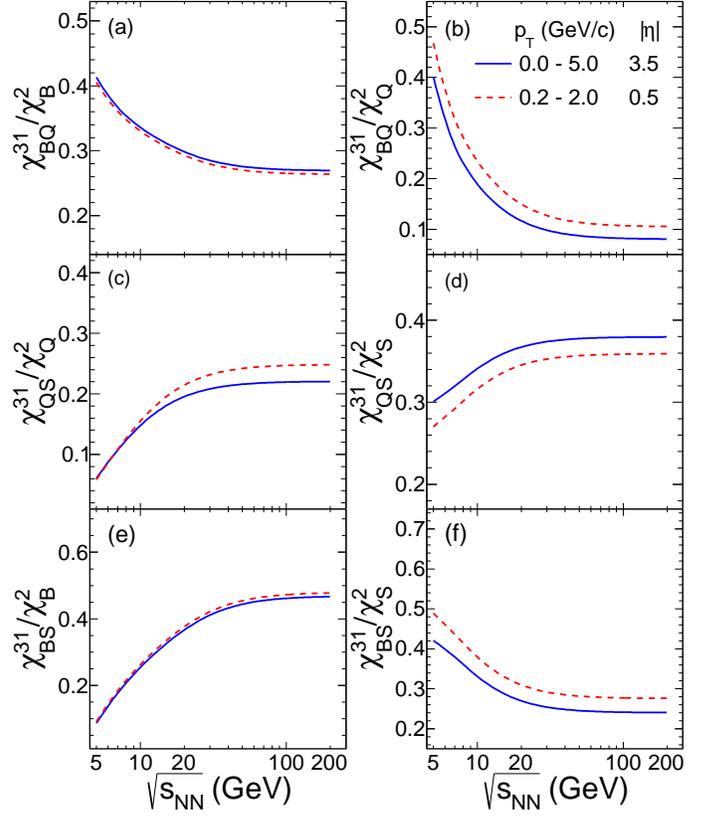}
 \caption{Ratios of off-diagonal susceptibility to second order susceptibility 
($\chi_{xy}^{31}/\chi_{x,~y}^2$) as a function of \sqsn for full $p_T$ 
and $|\eta|$ acceptance, and within 0.2 $\le p_T$ GeV/$c \le$ 2.0 with $|\eta| 
\le$ 0.5 from the HRG model, where $x$ and $y$ stand for either $B$, $Q$ 
or $S$. Panels (a) and (b) shows the off-diagonal susceptibility ratios for 
charged baryons. Panels (c) and (d) shows the off-diagonal susceptibility 
ratios for charged particles with at least one strange quark. Panels (e) 
and (f) shows the off-diagonal susceptibility ratios for baryons with strange 
quark.}
 \label{fig:netBQ31_pt}
 \end{center}
 \end{figure}

For completeness, we have also studied the ratios of $\chi^{31}/\chi^2$ as 
a function of collision energies. Figure~\ref{fig:netBQ31_pt} shows the \sqsn 
dependence of $\chi_{QS}^{31}/\chi_Q^2$, $\chi_{BQ}^{31}/\chi_Q^2$, 
$\chi_{BQ}^{31}/\chi_B^2$, $\chi_{QS}^{31}/\chi_S^2$, 
$\chi_{BS}^{31}/\chi_B^2$, and $\chi_{BS}^{31}/\chi_S^2$ for two different 
$p_T$ and $\eta$ acceptances. These susceptibility ratios are different from 
the $\chi^{13}/\chi^2$ ratios since we take the third derivative of the first 
conserved quantity. The effect will be more if the first considered conserved 
number is electric charge or strangeness.

\subsection{Effect of different conserved charged states}
\begin{figure}[h]
 \begin{center}
\includegraphics[width=0.47\textwidth]{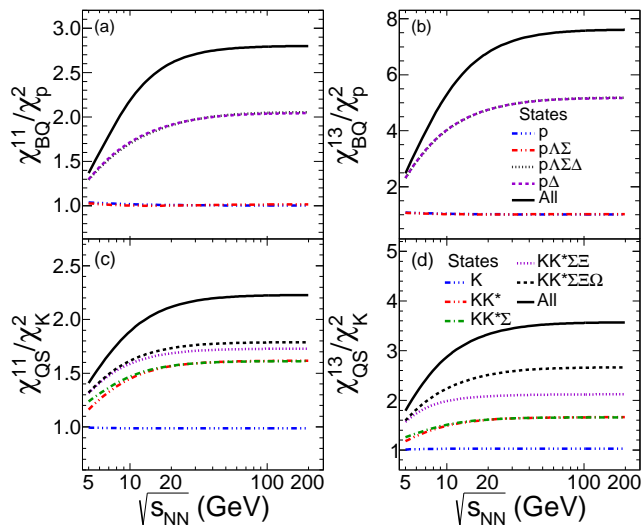}
 \caption{The variation of $\chi_{BQ}^{11,13}/\chi_{p}^2$ and 
$\chi_{QS}^{11,13}/\chi_{K}^2$ as a function of \sqsn within 0.2 $\le p_T$ 
(GeV/$c$) $\le$ 2.0 and $|\eta| \le$ 0.5. Panels (a) and (b) shows the 
susceptibility ratios for charged baryon and panels (c) and (d) shows the 
susceptibility ratios charged particles with strangeness content. Also shown 
are the comparison of effect of inclusion of particles with higher charge or 
strangeness states to the susceptibility ratios. The proton and kaon 
susceptibilities in the denominator are from the primordial production.}
 \label{fig:netBQ13_states}
 \end{center}
 \end{figure}
\begin{figure}[h]
 \begin{center}
\includegraphics[width=0.47\textwidth]{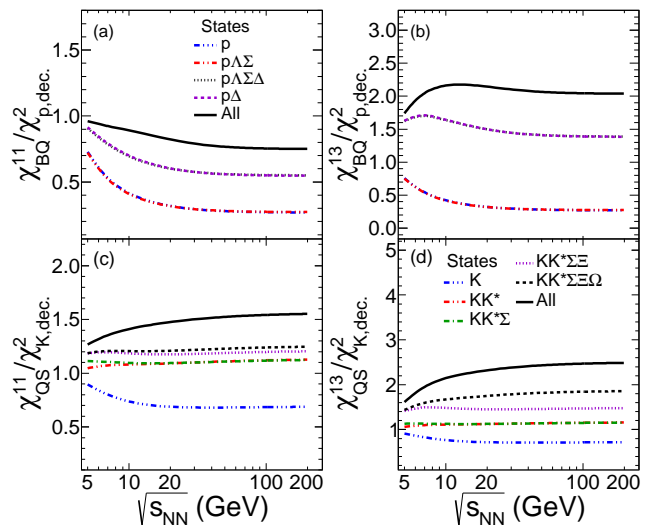}
 \caption{The variation of $\chi_{BQ}^{11,13}/\chi_{p}^2$ and 
$\chi_{QS}^{11,13}/\chi_{K}^2$ as a function of \sqsn within 0.2 $\le p_T$ 
(GeV/$c$) $\le$ 2.0 and $|\eta| \le$ 0.5. Panels (a) and (b) shows the 
susceptibility ratios for charged baryon and panels (c) and (d) shows the 
susceptibility ratios charged particles with strangeness content. Also shown 
are the comparison of effect of inclusion of particles with higher charge or 
strangeness states to the susceptibility ratios. The proton and kaon 
susceptibilities in the denominator are from the primordial as well as 
resonance decay contributions.}
 \label{fig:netBQ13_states_decay}
 \end{center}
 \end{figure}
Experimentally it is difficult to measure all the baryons or strange particles 
produced in the collisions. Hence, we have not considered the correlated 
susceptibilities related to strange baryons ($\chi_{BS}$).  Further, in the 
denominator we need to take the susceptibilities of $\chi_p$ and $\chi_K$  for 
protons (anti-protons) and kaons (anti-kaons), respectively, which may act as a 
proxy for baryons and strange particles. Figure~\ref{fig:netBQ13_states} shows 
the variation of $\chi_{BQ}^{11}/\chi_p^2$, $\chi_{BQ}^{13}/\chi_p^2$, 
$\chi_{QS}^{11}/\chi_K^2$ and $\chi_{QS}^{13}/\chi_K^2$ ratios as a function of 
\sqsn for various charge states. In case of baryon charge correlations 
($\chi_{BQ}$) only charged baryons (anti-baryons) contribute, starting from 
proton to higher mass charged baryons. For each of the cases, the $\chi_{BQ}$ 
are compared with the inclusion of higher charge states. There is significant 
difference in $\chi_{BQ}^{11}/\chi_p^2$ and $\chi_{BQ}^{13}/\chi_p^2$ after 
inclusion of higher charge states like $\Delta^{++}$. Since all the baryons have 
baryon number 1, only the baryons with higher electric charge number contribute 
to the increase of baryon charge correlations. Similarly, for strange electric 
charge correlations ($\chi_{QS}$) only charged strange particles contribute 
starting from kaon and higher mass charged strange particles. There is a strong 
dependence of $\chi_{QS}^{11}/\chi_K^2$ and $\chi_{QS}^{13}/\chi_K^2$ on the 
inclusion of higher order strange charged states. 

\begin{figure}[ht]
 \begin{center}
\includegraphics[width=0.45\textwidth]{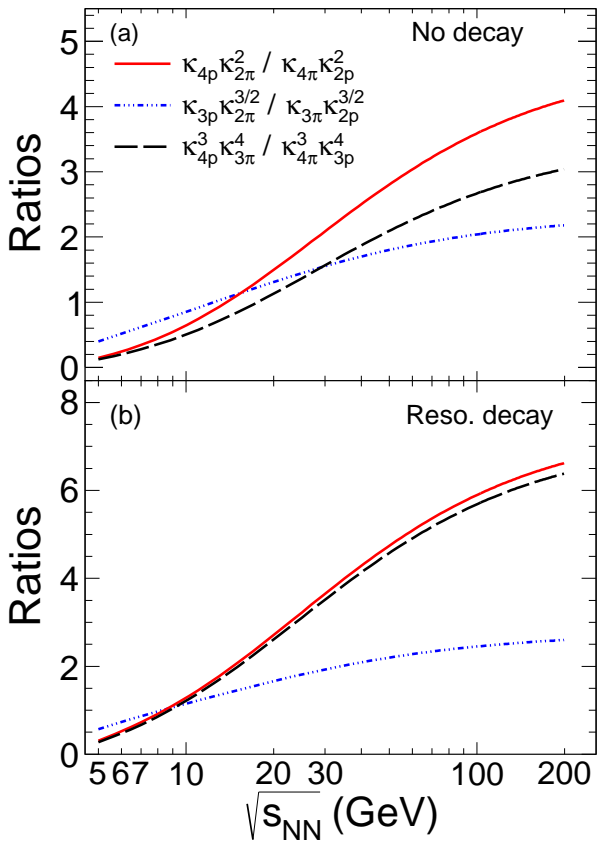}
 \caption{Various cumulant ratios of proton and pion as a functions of \sqsn 
within 0.2 $\le p_T$ GeV/$c \le$ 2.0 and $|\eta| \le$ 0.5. The upper panel 
(a) shows for the primordial proton and pion production, the lower panel (b) 
shows the proton and pion production from primordial and resonance decay 
contributions.}
 \label{fig:mixedratios_ppi}
 \end{center}
 \end{figure}

The measured proton (anti-proton) and kaon (anti-kaon) have contributions 
from primordial production as well as from resonance decay. We have studied the 
resonance decay effect on the mixed susceptibility ratios by taking the 
resonance contribution in the denominator ($\chi_p$ and $\chi_K$). The 
resonance decay contributions to the proton and kaon susceptibilities are 
estimated as discussed in~\cite{Mishra:2016qyj,Nahrgang:2014fza}. 
Figure~\ref{fig:netBQ13_states_decay} shows the collision energy dependence of 
$\chi_{BQ}^{11}/\chi_p^2$, $\chi_{BQ}^{13}/\chi_p^2$, $\chi_{QS}^{11}/\chi_K^2$ 
and $\chi_{QS}^{13}/\chi_K^2$ ratios for different charged states. The 
denominators $\chi^2_{p}$ and $\chi^2_K$ have contributions from primordial 
as well as resonance decay. The numerators have contributions from primordial 
particles and without decay of resonances. The $\chi^{11}_{BQ}/\chi^2_p$ in 
Fig.~\ref{fig:netBQ13_states} and Fig.~\ref{fig:netBQ13_states_decay} show 
different energy dependence because of the resonance decay contributions to the 
protons in the denominator for Fig.~\ref{fig:netBQ13_states_decay}. Further, 
the baryon-charge correlation $\chi_{BQ}^{11}/\chi_p^2$ and 
$\chi_{BQ}^{13}/\chi_p^2$ (case all) in Fig.~\ref{fig:netBQ13_states} show 
different energy dependence compared to the $\chi_{BQ}^{11}/\chi_B^2$ in 
Fig.~\ref{fig:netBQ11_eta} and $\chi_{BQ}^{13}/\chi_B^2$ in 
Fig.~\ref{fig:netBQ13_eta}. This difference in the ratios arise due to the 
difference in $\chi^2$ for protons and baryons. The $\chi^2$ of the baryons is 
always higher than the $\chi^2$ of protons and also has an energy dependence.

\section{Baseline for critical fluctuations}
\label{sec:baselineratios}
In order to locate the critical point in the QCD phase diagram, one may compare 
the experimentally measured cumulants of multiplicity distributions with the 
theoretical model predictions. These theoretical models, without any phase 
transition (non-CEP), may provide a baseline contribution to the measured 
cumulants from data. As proposed in~\cite{Athanasiou:2010kw}, there are 
different combinations of cumulants of protons and pions which are independent 
of QCD based coupling parameters. Some of such cumulant ratios are : 
$(\kappa_{3p}\kappa_{2\pi}^{3/2})/(\kappa_{3\pi}\kappa_{2p}^{3/2})$, 
$(\kappa_{4p}\kappa_{2\pi}^2)/(\kappa_{4\pi}\kappa_{2p}^2)$, 
$(\kappa_{4p}^3\kappa_{3\pi}^4)/(\kappa_{4\pi}^3\kappa_{3p}^4)$, 
$\kappa_{2p2\pi}^2/(\kappa_{4\pi}\kappa_{4p})$ and 
$\kappa_{2p1\pi}^3/(\kappa_{3p}^2\kappa_{3\pi})$.
Where, $\kappa_{nx}$ are the cumulants of different order with $n=1, 2, 3, 4$ 
and $x$ being the proton or pion. In case of non-interacting particles in the 
classical Boltzmann regime, in the absence of critical fluctuations, the 
fluctuations of particle number follow Poisson statistics. 
Hence, after subtracting the Poisson contribution from the above cumulant 
ratios, one can predict the contributions of critical fluctuations in the 
cumulant ratios. These subtracted cumulant ratios will be at 1 in the proximity 
of the critical point. However, keeping in mind that the experimental 
measurements are performed within a finite kinematic acceptance and can accept 
only a fraction of particles, the HRG model based baseline will be a better 
substitute compared to Poisson expectations. We have estimated the baseline 
for the above cumulant ratios using the HRG model, which will give the thermal 
contribution to the cumulant ratios. The first two ratios are the ratios of the 
skewness and kurtosis of protons and pions, where skewness 
$=\kappa_3/\kappa_2^{3/2}$ and kurtosis $=\kappa_4/\kappa_2^2$. The last two 
ratios involve mixed cumulants of protons and pions. In the HRG model, there is 
no interaction between the particles, hence the contribution of mixed cumulants 
will be zero. Figure~\ref{fig:mixedratios_ppi} shows the first three ratios, 
defined above, as a function \sqsn. The upper panel of 
Fig.~\ref{fig:mixedratios_ppi} shows the different cumulant ratios by taking 
primordial pions and protons, the lower panel show the cumulant ratios of 
protons and pions which have contributions from both primordial as well as  
resonance decay. These ratios are calculated using the HRG model should and can 
be considered as baseline instead of Poisson contribution as baseline. 

\section{Summary}
\label{sec:summary}
In summary, keeping in mind the importance of mixed susceptibilities
and pion-proton cumulants in unraveling the phase structure of QCD,
we have provided the variation of these observables as a function of
colliding beam energy within the ambit of a HRG model. The results
presented in this paper will provide the required baseline for the
corresponding measurements in the experiments. 

We have studied the effect of experimental acceptance on the ratios 
of the mixed susceptibilities using a hadron resonance gas model. We have 
considered $\chi_{BQ}^{11}/\chi_B^2$, $\chi_{BQ}^{11}/\chi_Q^2$, 
$\chi_{QS}^{11}/\chi_Q^2$, $\chi_{QS}^{11}/\chi_S^2$, 
$\chi_{BS}^{11}/\chi_B^2$ and $\chi_{BS}^{11}/\chi_S^2$ ratios, which have 
contributions from charged baryons (anti-baryons), charged strange or baryon 
strange particles.
There is a small dependence on $p_T$ and $\eta$ for the above susceptibility 
ratios. However, there is a strong energy dependence for all the ratios. These 
susceptibility ratios are very sensitive to the different charge states. 
In a real experimental situation, it is difficult to measure the strange 
baryons ($\Lambda^0$ and higher masses strange baryons) on an event-by-event 
basis and calculate their cumulants for fluctuations, specially at lower beam 
energies. Instead we have concentrated on measurable ratios like 
$\chi_{BQ}^{11}/\chi_p^2$, $\chi_{BQ}^{13}/\chi_p^2$, $\chi_{QS}^{11}/\chi_K^2$ 
and $\chi_{QS}^{13}/\chi_K^2$ and presented the HRG expectations for 
primordial and resonance decay contributions. 

We have also provided a HRG based calculation of realistic baseline (non-CEP 
expectation) for a set of new observables recently proposed for critical 
fluctuations using the different combinations of cumulant ratios of protons and 
pions~\cite{Athanasiou:2010kw}. The experimentally measured values of these 
ratios, after subtraction of the non-CEP contributions, are expected to be at 
unity in presence of CEP. These cumulant ratios of proton and pion from the HRG 
model calculations are observed to increase as a function of \sqsn. These ratios 
calculated using the HRG model should be used as baseline instead of Poisson 
expectation for the interpretation of data related to QCD critical point.

{\bf ACKNOWLEDGEMENT:}
BM acknowledges support from DST, DAE and SERB projects.


\end{document}